\documentclass[12pt, fleqn]{article}
\usepackage[round]{natbib}
\usepackage{amsmath}
\usepackage{amsfonts}
\usepackage{graphicx}
\usepackage[bookmarks=true]{hyperref}
\usepackage{rotating}%
\usepackage[nodisplayskipstretch]{setspace}
\usepackage{comment} 
\usepackage{threeparttable}
\usepackage{subfigure}

\newtheorem{theorem}{Theorem}

\author{  Ivan Medovikov\footnote{Brock University, St Catharines ON L2S 3A1, Canada; email: \textsf{imedovikov@brocku.ca} }  \qquad Valentyn Panchenko\footnote{Economics, UNSW Business School, Sydney NSW 2052, Australia;  email: \mbox{\textsf{valentyn.panchenko@unsw.edu.au}}} \qquad Artem Prokhorov\footnote{University of Sydney Business School; Sydney NSW 2006, Australia; email: \textsf{artem.prokhorov@sydney.edu.au}  } \footnote{CEBA, St.Petersburg State University, Russia} \footnote{CIREQ, University of Montreal, Canada}}

\title{\textbf{Efficient estimation of parameters in marginals in semiparametric multivariate models\footnote{Helpful comments of seminar participants at University of Toronto, University of Pittsburgh, UNSW Sydney, Concordia University, QMF, International Panel Data and FESAMES are gratefully acknowledged. Prokhorov's research was supported by a grant from Russian Science Foundation (20-18-00365).}}}

\date{
{\small
August 2023}\\\vspace{1cm}
}

\begin{document}

\maketitle
\begin{abstract}

We consider a general multivariate model where univariate marginal distributions are known up to a 
parameter vector and we are interested in estimating that parameter vector without specifying the joint distribution, except for the marginals. If we assume independence between the marginals and maximize the resulting quasi-likelihood, we obtain a consistent but inefficient QMLE estimator.  If we assume a parametric copula (other than independence) we obtain a full MLE, which is efficient but only under a correct copula specification and may be biased if the copula is misspecified. Instead we propose a sieve MLE estimator (SMLE) which improves over QMLE but does not have the drawbacks of full MLE. We model the unknown part of the joint distribution using the Bernstein-Kantorovich polynomial copula and assess the resulting improvement over QMLE and over misspecified FMLE in terms of relative efficiency and robustness.  We derive the asymptotic distribution of the new estimator and show that it reaches the relevant semiparametric efficiency bound. Simulations suggest that the sieve MLE can be almost as efficient as FMLE relative to QMLE provided there is enough dependence between the  marginals. 
We demonstrate practical value of the new estimator with 
financial risk management examples, where 
the use of SMLE leads to superior Value-at-Risk predictions. The paper comes with supplementary materials which include all proofs, codes and datasets, details of implementation and an additional application to insurance claims. 




\end{abstract}

\thispagestyle{empty}
\newpage
\setstretch{2}

\section{Introduction}\label{sec:introduction}

Consider an $m$-variate random variable $Y$ with joint pdf
$h(y_1,\ldots, y_m)$.  Let $f_1(y_1; \beta_1), \ldots,$ $f_m(y_m; \beta_m)$ denote the
corresponding marginal pdf's, known up to a parameter vector. 
The dependence structure between the marginals is not parameterized.  We
observe an i.i.d.~sample
$\{\mathbf{y}_i\}_{i=1}^N=\{y_{1i},\ldots,y_{mi}\}_{i=1}^N$ and we
are interested in estimating $\beta$ efficiently without assuming
anything about the joint distribution except for the marginals.

The literature on semiparametric copula models has focused on the
case when the marginals are specified nonparametrically and
the copula function is given a parametric form \cite[see, e.g.,][]{chen/fan/tsyrennikov:06, segers/akker/werker:08}, which is an appropriate setting for some financial applications where it is important to parameterize dependence.  In our setting, dependence is used solely to provide more precision in estimating marginal parameters, so we study the converse problem.

Our starting point is the marginals known up to a parameter vector. However this does not preclude some misspecification in the marginals. In particular, our results still hold in cases when misspecification does not lead to inconsistency of estimation for the feature of interest. Because we deal with generic likelihoods, in essence we allow for the marginals to be incorrect as long as they have zero-mean score functions at the pseudo-true parameter \citep{white1982maximum}.

We will use the well known representation of log-joint-density  in the marginal and copula density:
\begin{equation}\label{eq:density_decomposition}
\ln h(y_1,\ldots, y_m; \beta) = \sum_{j=1}^m\ln f_j(y_j;\beta_j) + \ln c(F_1(y_1;\beta_1),\ldots, F_m(y_m;\beta_m)),
\end{equation}
where $c(\cdot)$ is a copula density, $F_j(\cdot)$ denotes the
corresponding marginal cdf and where we collect all parameters of the marginals in one vector $\beta$ but allow for each marginal to depend on distinct subvectors of $\beta$. 
Sklar's
(\citeyear{sklar:59}) theorem which states that any continuous
joint distribution can be represented by
a unique copula function of the marginal cdf's. This is valid for any $m$  but in simulations and applications we focus on the values $m=\{2,3\}$ due to the curse of dimensionality. 

It is well understood that the parameters of the marginals can be
consistently estimated by maximizing the likelihood under the assumption of independence between the marginals -- this is the so called
quasi maximum likelihood estimator, or QMLE. The copula term in (\ref{eq:density_decomposition}) is zero in this case because the independence copula density is one.  However, QMLE ignores the dependence information and is not
efficient if marginals are not independent. For highly dependent marginals, the efficiency loss relative to the correctly specified full likelihood is quite large. \cite{joe:05}, for instance, reports up to 93\% improvements in relative efficiency over QMLE in simulations when a correctly specified full likelihood is used. We note that the marginals do not have to have common parameters for this result to apply.\footnote{Under dependence, random variables are constrained by an additional functional relationship that typically helps to identify the parameters of the marginals. Specifically, in the bivariate case, ${y_1=F_1^{-1}(C^{-1}(\eta|F_2(y_2;\beta_2));\beta_1)}$, where $C^{-1}(\eta|F_2(y_2;\beta_2))$ is the generalized inverse of conditional copula $C(u_1|u_2)=\tfrac{\partial  C(u_1,u_2)}{\partial u_2}$ and $\eta \in [0,1]$ is a uniform random variable. See some examples with specific marginals for the simpler cases of extreme dependence in Section~\ref{sec:simulations}.}  

There are numerous estimators that improve on QMLE and remain robust to dependence. 
For example, \cite{prokhorov/schmidt:09b} propose stacking the score functions from the marginal distributions and applying the Generalized Method of Moments (GMM) machinery to achieve the improvements via the use of correlation between the marginal scores\footnote{The GMM is also known as the GEE method, the method of generalized estimating equations \cite[see, e.g.,][]{hansen:82, Godambe/thompson:78}}; \cite{nikoloulopoulos/joe/chaganty:11} use a similar approach and construct a weighted sum of the marginal scores by fitting and discretizing a multivariate normal model for the scores. These estimators are simple because they are based on linear combinations of the marginal scores, but they 
cannot attain full efficiency unless the use of the true copula cannot improve upon the use of a linear combination of marginal scores.\footnote{We discuss the properties of these estimators in more details in Appendix~A.} 

The situation when using copula terms in the likelihood does not improve asymptotic efficiency over QMLE is known as copula redundancy.  \cite{prokhorov/schmidt:09b}
derived a necessary and sufficient condition for copula redundancy and showed that such situations are very rare. \cite{hao/etal:18} proposed a test of copula redundancy. Essentially, a parametric copula is redundant for the estimation of parameters in the marginals if and only if the copula score with respect to these parameters can be written as a linear combination of the marginal scores -- a condition generally violated for most commonly used parametric copula families and marginal distributions. 
As a consequence, significant efficiency gains due to the nonlinearity of the copula score in terms of the marginal scores  remain unexploited.

An obvious alternative that is more efficient asymptotically is a fully parametric estimation of the entire multivariate distribution by full MLE.  This means assuming a parametric copula specification in addition to the marginal distributions. It is now well understood that, unlike QMLE, FMLE is generally not robust to copula misspecification.  That is, the efficiency gains will come at the expense of an asymptotic bias if the joint density is misspecified.  \cite{prokhorov/schmidt:09b} point out that there are robust parametric copulas, for which the pseudo MLE (PMLE) using an incorrectly specified copula family leads to a consistent estimation.  However, copula robustness is problem specific and some robust copulas are robust because they are redundant.  So finding a general class of robust non-redundant copulas remains an unresolved problem.

In this paper we 
investigate whether we can obtain a consistent estimator of $\beta$ that is relatively more efficient than QMLE, by
modelling the copula term nonparametrically.  We use sieve MLE (SMLE) to do that. The questions we ask
are whether a sieve-based copula approximator can provide the robust non-redundant alternative to the QMLE and PMLE, and what is the semiparametric efficiency bound for the SMLE of $\beta$.  So our paper relates to the literature on sieve estimation \cite[see,
e.g.,][]{ai/chen:03, newey/powell:03, bierens:14} and on semiparametric efficiency bounds \cite[see,
e.g.,][]{severini/tripathi:01, newey:90}, including bounds for rank-based copula estimators \cite[see, e.g., ][]{segers/etal:14, hoff/etal:14}. The paper is similar to \cite{hu/etal:17} in that they also use a sieve MLE involving the Bernstein polynomial and discuss convergence and efficiency.\footnote{We thank an anonymous referee for pointing out the existence of this paper to us.} However, they work with copula functions, not copula densities, which complicates monotonicity restrictions, and their setup is restricted to  proportional hazard models; they do not discuss relative efficiency of SMLE, do not go beyond two dimensions or derive the Riesz representer.

The paper is organized as follows.  In Section 2 we define our estimator and prove consistency, asymptotic normality and semiparametric efficiency. 
Section 3
contains simulation results, confirming the significant efficiency gains permitted by SMLE.  Section 4 presents 
a financial application in two and three dimensions. Section 5 contains concluding remarks.\goodbreak

\section{Sieve MLE}
\label{section::sieve_mle}

Denote the true copula density by $c_o(\mathbf{u})$,
$\mathbf{u}=(u_1,\ldots,u_m)$, and denote the true parameter
vector by $\beta_o$.  Let $\beta_o$
belong to finite dimensional space $B \subset R^p$ and $c_o(\mathbf{u})$ belong to an
infinite-dimensional space \mbox{$\Gamma=\{c(\mathbf{u}): [0,1]^m
\rightarrow [0,1], \int_{[0,1]^m} c(\mathbf{u}) d\mathbf{u}=1, \int_{[0,1]^{m-1}} c(\mathbf{u})d\mathbf{u}_{\textrm{-} \ell}=1,\forall \ell \},$} where  $\mathbf{u}_{\textrm{-}\ell}$ excludes $u_\ell$. These conditions reflect the fact that any copula is a joint probability distribution on the unit cube $[0,1]^m$ with uniform marginals. Given a finite amount of data,
optimization over the infinite-dimensional space  $\Gamma$ is not
feasible. The method of sieves is useful for overcoming this problem. See Appendix~B for the basics of sieve MLE. 

Let $\Gamma_N$ denote a sequence of approximating spaces, called
sieves, such that $\bigcup_N\Gamma_N$ is dense in $\Gamma$.
One of the challenges of SMLE in our setting 
is ensuring that  $\Gamma_N$ 
consists of proper copula pdfs, that is, non-negative functions that integrate to one and have uniform marginals. Exponential or quadratic transformations are often used to ensure positivity and division by a normalizing constant is used to ensure that the sieve integrates to one \cite[see, e.g.,][]{chen/fan/tsyrennikov:06}. However, it is difficult to find an appropriate normalisation to ensure that all marginals are uniform. For example,  \cite{Anderson/prokhorov/zhu:21} show that very few of the popular nonparametric copula estimators satisfy this property in finite samples. Moreover, properties of normalised objects, namely, rates of convergence, may differ from those of the original sieve and may not be easy to derive.  A sieve which does not require any transformation to satisfy the proper copula conditions and has meaningful parameters is the Bernstein-Kantorovich polynomial \cite[see, e.g.,][]{sancetta:satchell:04}.  

\subsection{Bernstein-Kantorovich Sieve}\label{BersteinSieve}

The Bernstein-Kantorovich sieve is a tensor product sieve which uses beta-densities as basis functions; it can be written as follows:
\begin{eqnarray}\label{eq:bernstein-kantorovich-sieve}
c_{J_N}(\mathbf{u})=(J_N)^m \sum_{v_1=0}^{J_N-1}\dots \sum_{v_m=0}^{J_{N}-1}\omega_{\mathbf{v}}\prod_{l=1}^m
\left(\begin{array}{c}
  J_N-1 \\
  v_l
\end{array}\right)
u_l^{v_l}(1-u_l)^{J_N-v_l-1},
\end{eqnarray}
where $\omega_{\mathbf{v}}$ denotes parameters of the polynomial indexed by multi-index 
$\mathbf{v}=(v_1,\ldots,v_m)$ such that $0\leq \omega_{\mathbf{v}} \leq 1$ and
$\sum_{v_1=0}^{J_N-1}\dots \sum_{v_m=0}^{J_{N}-1}\omega_{\mathbf{v}}=1$. These restrictions ensure that the above equation is a proper density. The interpretation of the coefficients $\omega_{\mathbf{v}}$ is that they are probability masses on an $J_N \times \dots \times J_N$ grid \cite[see, e.g.,][]{ Zheng:11, burda/prokhorov:14}.\footnote{For simplicity we assume that $J_N$ is the same in each dimension $\ell$, but this assumption can be easily relaxed in cases where such asymmetry is required.}
In order to ensure that $c_{J_N}(\mathbf{u})$ is a copula density, i.e. that its marginals are uniform, we further require that  $\sum_{\mathbf{v}_{\textrm{-}\ell|v_\ell}} \omega_{\mathbf{v}}=1/J_N$, where multiple summations are performed over all elements of $\mathbf{v}$ 
except $v_\ell,\, \ell=1,\dots, m$ for each fixed value of $v_\ell$, where $v_\ell=0,\dots J_N-1$. Hence, there are $J_N\times m$ of these restrictions in total.

The weights $\omega_{\mathbf{v}}$ are akin to a multivariate empirical copula density values,
$\omega_{\mathbf{v}} = \frac{1}{N}\sum_{i=1}^N\mathbb{I}(u_i\in H_{\mathbf{v}})$,
where $u_i=(u_{i1}, \ldots, u_{im}) \in [0, 1]^m$, $\mathbb{I}(\cdot)$ is the indicator function and
\begin{eqnarray}
H_{\mathbf{v}}=\left[\frac{v_1}{J_N},\frac{v_1+1}{J_N}\right]\times\dots\times
\left[\frac{v_m}{J_N},\frac{v_m+1}{J_N}\right].
\label{eq:hist}
\end{eqnarray}
With these $\omega_{\mathbf{v}}$'s, the Bernstein-Kantorovich polynomial sieve can be viewed as a smoothed copula histogram where smoothing is done by the product of beta-densities.  Alternatively, it can be viewed as a mixture of a product of beta-densities in $u$ with mixing weights $\omega_{\mathbf{v}}$ \cite[see, e.g.,][]{burda/prokhorov:14}.

\cite{sancetta:07} derives the rates of
convergence of the Bernstein-Kantorovich copula to the true copula. \cite{hmp:20} explore weak and strong uniform convergence of beta kernels on expanding compact sets on $(0,1)$. \cite{Petrone:Wasserman:02} and \cite{burda/prokhorov:14} establish consistency of the Bernstein-Kantorovich polynomial when used as a prior on the space of densities on $[0,1]^m$. 
\cite{ghosal:01} and references therein discuss the rate of
convergence of the sieve MLE based on the Bernstein polynomial (only
for one-dimensional densities). Uniform approximation results for the univariate and
bivariate Bernstein density estimator can be also found in \cite{Vitale:75} and 
\cite{Tenbusch:94}.  As $J_N\rightarrow \infty,$ $c_{J_N}(\mathbf{u})$ is known to converge to the probability limit of the empirical copula density estimator at every point on $[0,1]^{m}$ where the limit exists,
and if it is continuous and bounded then the convergence is uniform \citep[see, e.g.,][]{Lorentz:86}.\footnote{In practice, the choice of $J_N$ is important to the extent to which it affects the bias-variance trade-off in finite samples: as shown by \cite{sancetta:satchell:04}, the Bernstein-Kantorovich sieve has bias of the order $O(J_N^{-1})$, which is the same as for a histogram or kernel-smoothers, but variance of order $O(J_N^{m/2})$ inside the hypercube, which is a square-root of the rate for a histogram or kernel estimator. The theoretically optimal order for $J_N$ in the MSE sense is $O(N^{\frac{2}{m+4}})$, which is greater than for standard nonparametric estimators such a histogram or first-order kernels, implying relatively little smoothing required for this sieve. Suboptimal growth of $J_N$ affects the balance of bias and variance but has no effect on semiparametric efficiency of $\hat{\beta}$ as long as $J_N \rightarrow \infty,\, \frac{J_N}{N}\rightarrow 0$ and Assumption A4 holds.} 


This sieve is particularly attractive in our setting because of the uniform rate of convergence results available for $c_{J_N}$ and because of the empirical copula density interpretation of $\omega_{\mathbf{v}}$.  The former ensures a relatively fast convergence compared to other tensor product sieves, which we observe in simulations, while the latter permits natural adaptive dimension reduction based on dropping $\omega_\mathbf{v}$'s which correspond to sparsely populated grid cells. Other potential explanations for the good performance of the sieve in economic, finance, actuarial and risk management tasks are that such data have inhomogeneous dependence structures, are not highly correlated and sparse \cite[see, e.g.,][]{diers/etal:12}. 

\subsection{Asymptotic Properties}

Let the sieve for $\Theta=B \times \Gamma$ be denoted by $\Theta_N=B\times \Gamma_N$, where $\Gamma_N$ contains a generic vector of copula parameters $\gamma$, and let $\theta=(\beta', \gamma)$. For the special case of the Bernstein-Kantorovich copula, $\gamma = \omega_\mathbf{v}$.  

We now list identification and smoothness assumptions. Versions of these are commonly used in sieve estimation literature \cite[see, e.g.,][]{shen:97, ai/chen:03,  chen/fan/tsyrennikov:06,  chen:07, bierens:14}. In the discussion of these assumptions we focus on what is new to our copula-based settings. 

\bigskip
\noindent\textbf{Assumptions}\\
\textbf{A1} (identification) $\beta_o \in $ int$(B)\subset R^p$, $B$ is compact
and there exists a unique $\theta_o$ which maximizes $E [\ln
h(\mathbf{Y}_i;\theta)]$ over $\Theta=B \times \Gamma$.\\
\textbf{A2} (smoothness
)  \mbox{$\Gamma = \{c=\exp(g):g\in \Lambda^r([a,b]^m), \int c(\mathbf{u}) d\mathbf{u} = 1,\int_{[0,1]^{m-1}} c(\mathbf{u})d\mathbf{u}_{\textrm{-}\ell}=1,\,\forall \ell\},$} where $\Lambda^r([a,b]^m)$ denotes the H\"older class of $r$-smooth functions on  $[a,b]^m$, $\forall [a, b]\subset (0,1)$, $r >
1/2, $ and $\ln f_j(y_j;\beta), j=1,\ldots, m,$
are twice continuously differentiable w.r.t.~$\beta$.
\bigskip

The smoothness condition restricts log-copula-densities to the class of real-valued, continuously differentiable functions whose $J$-th order derivative
satisfies H\"older's condition inside the hypercube
\begin{equation}\nonumber 
|D^J g(x) - D^J g(y) | \le K|x-y|_E^{r-J}, \text{for all }x,y \in [a,b]^m \text{ and some }r \in (J, J+1]
\end{equation}
where $D^\alpha=\frac{\partial^\alpha}{\partial x^{\alpha_1}_1
\ldots \partial x^{\alpha_m}_m}$ is the derivative operator, $\alpha=\alpha_1+\ldots+\alpha_m$, $|x|_E = (x'x)^{1/2}$ is the
Euclidean norm and $K$ is a positive constant. We exclude the 
edges of $[0,1]^m$ where copula densities can be unbounded.\footnote{An alternative is to employ at the edges an expanding set sequence \cite[see, e.g.,][]{hmp:20} or a trimming or weighting scheme \cite[see, e.g.,][]{hmp:20, HILL201618}. We leave such approaches for future work.} Commonly used densities, including copula densities, belong to the H\"older class on $[a,b]^m$, and various linear sieves, as well as the Bernstein-Kantorovich polynomial sieve, are known to approximate such functions well. 
$\Lambda^r([a,b]^m)$ is one of the most popular function classes in nonparametric estimation literature \cite[see, e.g.,][]{horowitz:98, chen:07}.

In this semi-parametric settings, the initial parameter vector is infinite-dimensional because it contains the nonparametric part, $\ln c$, along with $\beta$. The asymptotic distribution of $\hat{\beta}$ -- the first $p$ elements of
$\hat{\theta}$ -- depends on the behavior of $\hat{\theta}$ as its dimension grows.  By the Gram\'er-Wold device, this distribution is
normal if, for any $\lambda \in R^p, \lVert \mathbf{\lambda}
\rVert \ne 0 $, the distribution of the linear combination $\lambda'
\hat{\beta}$ is normal. Note that $\lambda' \beta$ is a functional of
$\theta$, call it $\rho(\theta)$. Given a sieve
estimate $\hat{\theta}$, the asymptotic distribution of $\rho(\hat{\theta})$ depends on smoothness of the functional and on the convergence rate of the
nonparametric part of $\hat{\theta}$ \cite[see, e.g.,][]{shen:97}.
In our setting, the functional is simple and smooth.  But the rate of convergence of the nonparametric part of $\hat{\theta}$ may be quite slow especially if $m$ is large. It is a well established result in univariate settings that in such cases the smoothness of $\rho(\beta)$ compensates for this, and $\hat{\beta}$ achieves $\sqrt{N}$-convergence
\cite[see, e.g.,][]{bierens:14}. We obtain a similar result in multivariate settings.

Let $\dot{l}(\theta_o)[\nu]$ denote the directional
derivative, evaluated at $\theta_o$, of the log-likelihood in direction $\nu = (\nu_\beta',
\nu_\gamma)' \in V$, where $V$ is the linear span of $\Theta -
\{\theta_o\}$.  Then,
\begin{displaymath}
\begin{array}{rcl}
\dot{l}(\theta_o)[\nu] & \equiv & \lim_{t\rightarrow 0} \left.\frac{\ln h  (y, \theta+t\nu)- \ln h(y, \theta)}{t}\right|_{\theta=\theta_o}  =  \frac{\partial \ln h (y, \theta_o)}{\partial \theta'}[\nu]\\
& = & \sum_{j=1}^m \left\{ \frac{\partial \ln f_j(y_j, \beta_o)}{\partial \beta'} +\left(
\frac{1}{c (\mathbf{u})}
\left.\frac{\partial c (\mathbf{u})}{\partial u_j}\right)\right|_{u_k=F_k(y_k, \beta_o)} \frac{\partial F_j(y_j, \beta_o)}{\partial \beta'}  \right\} \nu_\beta +\left. \frac{1}{c (\mathbf{u})}
\nu_\gamma (\mathbf{u})\right|_{u_k=F_k(y_k, \beta_o)},
\end{array}
\end{displaymath}
where the last equation follows from (\ref{eq:density_decomposition}).
Similarly, define $\dot{\rho}(\theta_o)[\nu]$ as follows:
\begin{displaymath}
\begin{array}{rcl}
\dot{\rho}(\theta_o)[\nu] &\equiv& \lim_{t\rightarrow 0}
\left.\frac{\rho (\theta+t\nu)-
\rho(\theta)}{t}\right|_{\theta=\theta_o} =\lambda' \nu_\beta = \rho(\nu).
\end{array}
\end{displaymath}

Let $\langle \cdot, \cdot
\rangle$ denote the inner product based on the Fisher information metric on $V$ and let $||\cdot||$ denote the Fisher information norm on $V$. Then, $\langle \nu_1, \nu_2 \rangle =  E
\left[\dot{l}(\theta_o)[\nu_1]\dot{l}(\theta_o)[\nu_2] \right]$ and $||\nu|| = \sqrt{\langle \nu,
\nu \rangle}$, where expectation is with respect to the true
density $h$.  The closed linear span of $\Theta - \{\theta_o\}$
and the Fisher information metric form a
Hilbert space, call it $(\bar{V}, ||\cdot||)$.

Since $\rho(\theta)=\lambda' \beta$ is linear on $\bar{V}$, in order to show smoothness of
$\rho(\theta)$, we only need to establish that it is
bounded on $\bar{V}$, i.e. that $\sup_{ 0\ne \theta-\theta_o \in
\bar{V}} \frac{|\rho(\theta)-\rho(\theta_0)|}{||\theta-\theta_o||
}< \infty$.  Also, by the results in \cite{shen:97}, boundedness of $\rho(\theta)=\lambda' \beta$ is necessary for $\rho(\theta)=\lambda' \beta$ to be estimable at the $\sqrt{N}$-rate. 
Moreover, since $\dot{\rho}(\theta_o)[\nu] = \rho(\nu)$, boundedness of the directional derivative of $\rho(\theta)$ is equivalent to boundedness of $\rho(\theta)$ itself, i.e.~it is equivalent to  $\sup_{ 0\ne
\nu \in \bar{V}} \frac{|\dot{\rho}(\theta_o)[\nu]|}{||\nu|| }<
\infty$. Because $\rho(\nu) = \lambda'\nu_\beta$, this is the case if and only if $\sup_{\nu \ne 0, \nu\in
\bar{V}} \frac{|\lambda'\nu_\beta|^2}{||\nu||^2}<\infty$. So we
now show when this condition holds.

We follow \cite{ai/chen:03} and \cite{chen/fan/tsyrennikov:06} and look for the minimal componentwise Fisher  information matrix for $\beta$.  For our specific setting, this minimization problem can be written as follows: 
\begin{eqnarray} \label{eq:variance_problem}
\inf_{g_q} E\left[ \sum_{j=1}^m \left\{ \frac{\partial \ln
f_j(y_j, \beta_o)}{\partial \beta_q} + \left.\left(\frac{1}{c
(\mathbf{u})} \frac{\partial c (\mathbf{u})}{\partial
u_j}\right)\right|_{u_k=F_k(y_k, \beta_o)} \frac{\partial F_j(y_j,
\beta_o)}{\partial \beta_q} \right\} +
\left.\left(\frac{1}{c (\mathbf{u})}
g_q(\mathbf{u})\right)\right|_{u_k=F_k(y_k, \beta_o)} \right]^2,
\end{eqnarray}
where $E\left[\frac{1}{c(\mathbf{u})} g_q(\mathbf{u})\right]=0$.
Let $g^\ast_q$ denote the solution of (\ref{eq:variance_problem}), $q=1, \dots, p$, and let $g^\ast = (g_1^*, \ldots, g_p^*)$.  

We can now find the $\sup$ by writing
\begin{equation}\label{eq:sup}
\begin{array}{rcl}
\sup_{\nu \ne 0, \nu\in \bar{V}} \frac{|\lambda'\nu_\beta|^2}{||\nu||^2} & = & \sup_{\nu \ne 0, \nu\in \bar{V}} \left\{ |\lambda' \nu_\beta|^2 \left(E\left[ \dot{l}(\theta_o)[\nu]^2\right]\right)^{-1}\right\} =  \lambda' \left( E S_\beta S_\beta' \right)^{-1}\lambda,
\end{array}
\end{equation}
where
\begin{equation}
\begin{array}{rcl}
S_\beta' & = & \sum_{j=1}^m \left\{ \frac{\partial \ln f_j(y_j,
\beta_o)}{\partial \beta'} +\left( \frac{1}{c (\mathbf{u})}
\left.\frac{\partial c (\mathbf{u})}{\partial
u_j}\right)\right|_{u_k=F_k(y_k, \beta_o)} \frac{\partial F_j(y_j,
\beta_o)}{\partial \beta'}
 \right\} +\left.\left( \frac{1}{c (\mathbf{u})}
g^*(\mathbf{u})\right)\right|_{u_k=F_k(y_k, \beta_o)}, \\

g^\ast& =& (g^*_1, \ldots, g^*_p) \quad \text{and}\quad E\left[\frac{1}{c(\mathbf{u})} g^*_q(\mathbf{u})\right]=0.
\end{array}
\label{eq:score}
\end{equation}
Note that the second equality in (\ref{eq:sup}) is true because $g^\ast$ is the minimizer of $E\left[\dot{l}(\theta_o)[\nu]^2\right]$ over $\nu_\gamma$ at the true $\beta_0$. So $\rho(\theta)=\lambda' \beta$ is bounded if and only if $E S_\beta S_\beta'$ in (\ref{eq:sup}) is a finite
and positive definite matrix.  

\bigskip
\noindent \textbf{Assumption A3} (nonsingular information) Assume that $ES_\beta S_\beta'$ is finite and positive definite.

\bigskip

It is worth returning to the parametric setting to illustrate the intuition behind Assumption A3. In essence, $ES_\beta S_\beta'$ is the marginal Fisher information and Assumption A3 can be viewed as a non-redundancy condition of the true copula for the estimation of $\beta$ \cite[see][Section 4]{prokhorov/schmidt:09b}. Aside from technical failures such as moment non-existence, it assumes away cases when knowledge of the true copula cannot improve precision in the MLE of $\beta$ in principle. \cite{prokhorov/schmidt:09b} show that this 
happens only if the copula score is a linear combination of the marginal scores of $\beta$. For example, for a bivariate normal with a common mean and known correlation, the copula score for the mean is a linear combination of the marginal scores for the mean. However, this is not the case if the normal copula is replaced by the FGM copula or any other commonly used copula function with known dependence parameter \cite[see,][Examples 1, 5 and 6]{prokhorov/schmidt:09b}.

Having established smoothness of $\rho(\theta)$ we can use the Riesz representation theorem \cite[see,
e.g.,][p.~328]{kosorok:08} to derive the asymptotic distribution
of $\lambda'\beta$. Basically, the theorem states that for any
continuous linear functional $L(\nu)$ on a Hilbert space there
exists a vector $\nu^*$ (the Riesz representer of that functional)
such that, for any $\nu$, we have $L(\nu)=\langle \nu, \nu^* \rangle$, and the norm of the functional defined as
$||L||_* \equiv \sup_{||\nu|| \le 1 }||L(\nu)||$
is equal to $||\nu^*||$.  The representer will be used in the derivation of asymptotic
normality and semiparametric efficiency of the sieve MLE.

The Riesz representation theorem, when applied to $\dot{\rho}(\theta_o)[\nu]=\rho(\nu)$, suggests that there exists
a Riesz representer $\nu^* \in \bar{V}$  of $\rho(\nu)$, for which $\lambda'
(\hat{\beta}-\beta_o)=\langle \hat{\theta} -\theta_o, \nu^* \rangle$ and
$||\nu^*||=\sup_{||\nu||\le 1 }||\rho(\nu)||$. The first claim
implies that the distributions of $\hat{\beta}-\beta_o$
and of $\langle \hat{\theta} -\theta_o, \nu^* \rangle$ are identical, which is useful for proving asymptotic normality of $\sqrt{N}(\hat{\beta}-\beta_o)$.
The second claim is used in the proof of semiparametric efficiency.  Both of these claims are useful for deriving the explicit form of the representer for our settings.

It turns out we have already found $\nu^*$ when we showed smoothness of
$\rho(\theta)$ by finding $\sup_{\nu \ne 0, \nu\in \bar{V}}
\frac{|\lambda'\nu_\beta|^2}{||\nu||^2}$.  Since $\sup_{\nu \ne 0,
\nu\in \bar{V}}
\frac{|\lambda'\nu_\beta|^2}{||\nu||^2}=\sup_{||\nu||= 1
}||\rho(\nu)||^2$, the representer for our problem is a vector whose squared Fisher information norm is equal to $\sup_{\nu \ne 0, \nu\in \bar{V}}
\frac{|\lambda'\nu_\beta|^2}{||\nu||^2}=\lambda' \left( E S_\beta
S_\beta' \right)^{-1}\lambda$.  It is straightforward to show that this vector can be written as follows
\begin{equation}\label{eq:representer_nu}
\nu^* = \left(I, g^{*'} \right)'\left( E S_\beta S_\beta'
\right)^{-1}\lambda
\end{equation}
As a check we can see that the squared Fisher information norm of $\nu^*$ can be written as $||\nu^*||^2 = E\left[\dot{l}(\theta_o)[\nu^*]\dot{l}(\theta_o)[\nu^*] \right] =  \lambda'\left( ES_\beta S_\beta' \right)^{-1}\lambda$.

The last assumption required for asymptotic normality of $\sqrt{N}(\hat{\beta}-\beta_o)$ is an
assumption on the rate of convergence for the sieve MLE estimator
of the unknown copula function.  As in other sieve literature, we allow the sieve estimator to converge arbitrary
slowly -- smoothness of $\rho(\theta)$ compensates for that and
the parametric part of the estimator is still $\sqrt{N}$-estimable.  We also impose a boundedness condition on the second order term in the Taylor expansion of the sieve log-likelihood function.  This technical condition will usually follow from the smoothness assumption \textbf{A2} but we state it explicitly to simplify the proof.

\bigskip\noindent \textbf{Assumption A4} (convergence of sieve MLE and smoothness of higher order term in Taylor expansion) Assume (A) that $||\hat{\theta}-\theta_o ||=O_P(\delta_N)$ for
$(\delta_N)^{w} = o(N^{-1/2})$, $w>1$; (B) there exists $\Pi_N \nu^* \in
V_N - \{\theta_o\}$ such that $\delta_N ||\Pi_N \nu^* -
\nu^*||=o(N^{-1/2})$ and
(C) that, for any~${\theta:||\theta-\theta_o||=O_p(\delta_N)}$, the additional conditions on the second-order derivatives stated in Appendix C hold. 
\bigskip

A discussion of convergence rates of different sieves is provided by \cite{chen:07} and in references therein; general results on convergence rates of sieve MLE can be found in \cite{wong/severini:91, shen/wong:94}.  Basically, Assumption \textbf{A4}(A) covers all commonly encountered sieves.  For example, for the trigonometric sieve, \cite{shen/wong:94} show that its order of convergence is 
$O_p(N^{-r/(2r+1)})$, where $r$ is the H\"older exponent; 
for Bernstein-Kantorovich polynomial sieves, \cite{sancetta:satchell:04} show that its rate of convergence is 
$O_p(N^{-4/(m+4)})$ within the hypercube, where $m$ is the dimension; \cite{BOUEZMARNI:10} extend the results of \cite{sancetta:satchell:04} to $\alpha$-mixing data. Assumptions \textbf{A4}(B)-(C) are technical assumptions that control smoothness of the Riesz representer and  second order term in the expansion of the log-likelihood \cite[see][for details]{chen/fan:06}.

We can now state our main consistency and asymptotic efficiency results.

\begin{theorem}\label{th:CAN}
Under  \textbf{A1}-\textbf{A4},
$\sqrt{N}(\hat{\beta}-\beta_o ) \Rightarrow N(0,(E[S_\beta
S_\beta'])^{-1})$. 
\end{theorem}

\textbf{Proof.} All proofs are provided in Appendix C.


\begin{theorem}\label{th:bound}
Under  \textbf{A1}-\textbf{A4},
$||\nu^*||^2$ is the lower bound for semiparametric estimation of $\lambda'\beta$, i.e.~$\hat{\beta}$ is semiparametrically
efficient.
\end{theorem}

In practice, one needs to estimate the asymptotic variance in order to conduct inference on $\beta$. The matrix $E[S_\beta
S_\beta']$ can be estimated consistently as a sample average of $S_{\beta} S'_{\beta}$, once we obtain $\hat{\beta}$, $\hat{c}$, $\hat{g}^*_q$'s. Parameter estimates $\hat{\beta}$ and $\hat{c}$ are obtained in the sieve MLE but estimation of $g^*_q$ requires a separate sieve minimization problem.\footnote{An alternative estimator of $E[S_\beta
S_\beta']^{-1}$ was proposed by \cite{ackerberg/chen/hahn:12, ackerberg/chen/hahn/liao:14}. It uses the covariance matrix of all $p+ J_N^m$ model parameters (both parameters in the marginal and in the copula). The upper left $p \times p$ block of its inverse is used as the variance estimator.  However, this method assumes that the likelihood is separable in $\beta$ and $c$, which is not the case in our settings. This causes the estimate to be numerically unstable.} In our settings, consistent estimators of $g^*_q$, $q=1,\ldots,p$, are the solution to 
\begin{eqnarray} \label{eq:asvar}
 \arg \min_{g_q \in \mathbf{A}_N} \sum_{i=1}^N \left[ \sum_{j=1}^m \left\{ \frac{\partial \ln f_j(y_{ji}, \hat{\beta})}{\partial \beta_q} +\left(
\frac{1}{\hat{c} (\mathbf{\hat{u}}_i)}
\left.\frac{\partial \hat{c} (\hat{\mathbf{u}}_i)}{\partial u_j}\right)\right|_{\hat{u}_{ki}=F_k(y_{ki}, \hat{\beta})} \frac{\partial F_j(y_{ji}, \hat{\beta})}{\partial \beta_q}
 \right\}\left.+ \frac{1}{\hat{c} (\mathbf{\hat{u}}_i)} g_q(\mathbf{\hat{u}}_i)\right|_{\hat{u}_{ki}=F_k(y_{ki}, \hat{\beta})}\right]^2,
\end{eqnarray}
where $\mathbf{A}_N$ is one of the sieve spaces
discussed above and $\hat{\beta}$ and $\hat{c}$ are consistent estimates of $\beta$ and $c$ and $\int g_q(\mathbf{u})/\hat{c}(\mathbf{u})\, d\mathbf{u} =0$.  
See Appendix~D, for the practical aspects of our SMLE implementation using Bernstein-Kantorovich polynomials and Appendix~E for comparison with other tensor sieves.  

\section{Simulations}
\label{sec:simulations}


\begin{figure}[t]
	\centering
	\subfigure[]{\includegraphics[width=0.48\textwidth]{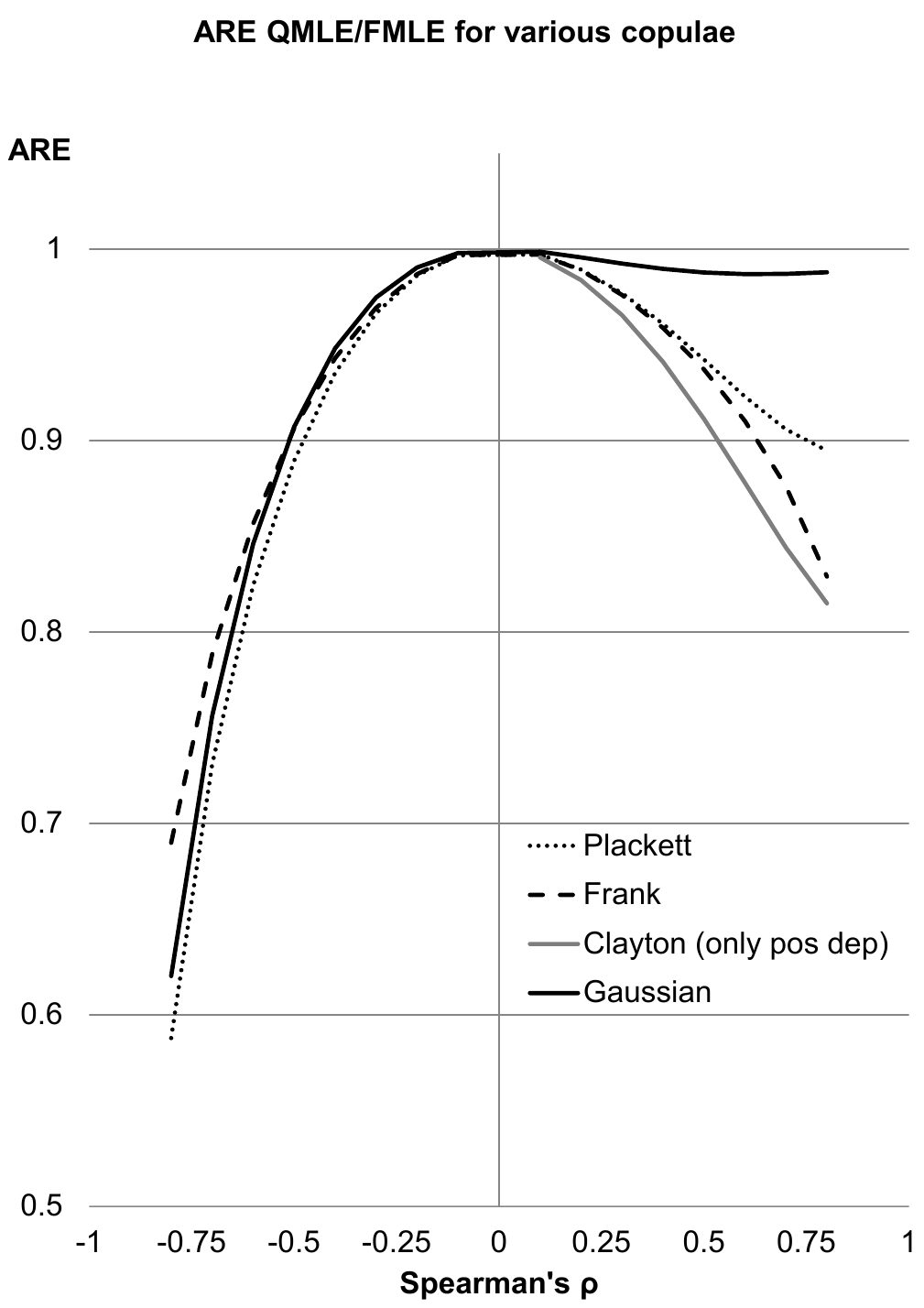}}\hfill
	\subfigure[]{\includegraphics[width=0.48\textwidth]{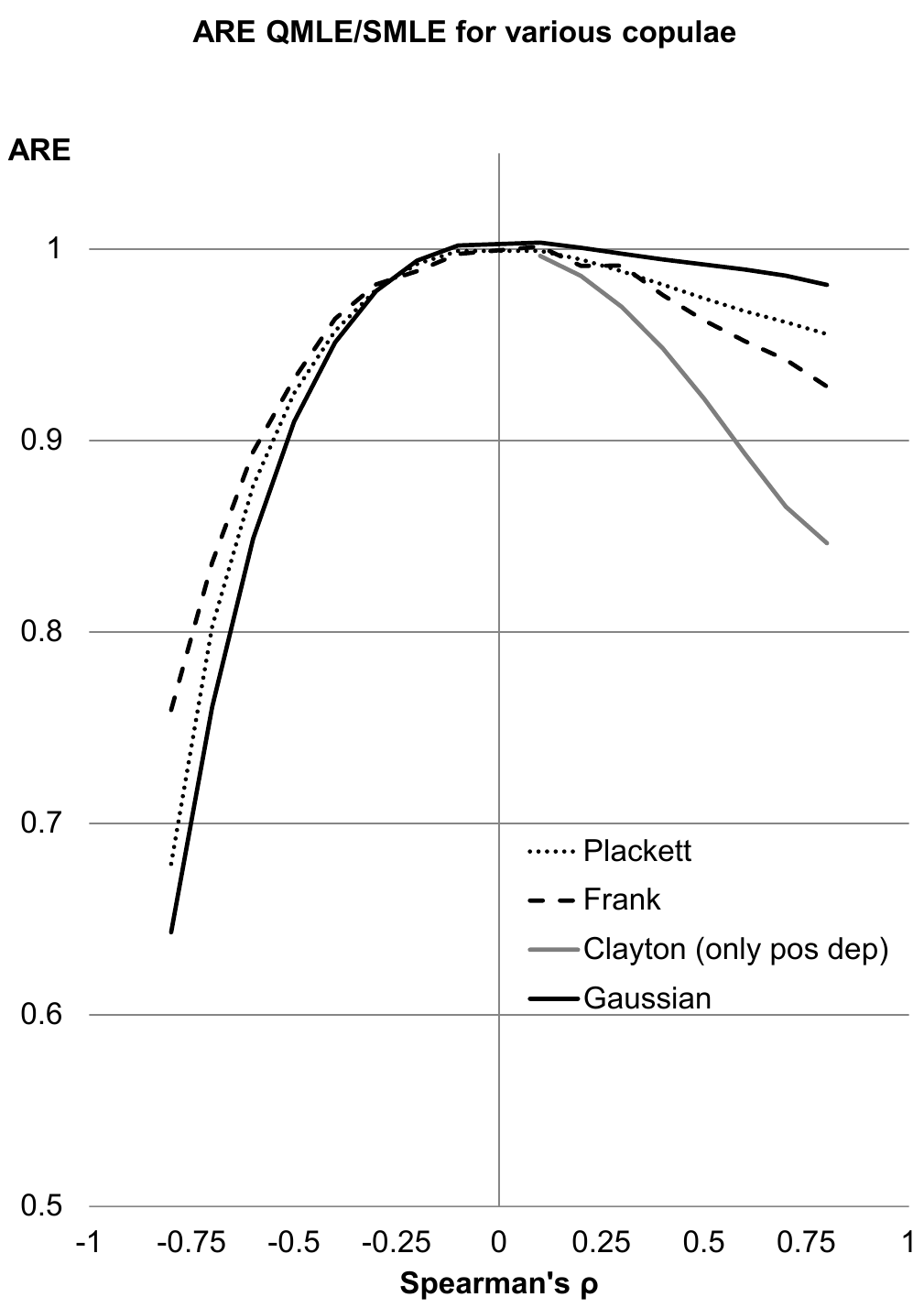}}
	\caption{Asymptotic relative efficiency of QMLE to FMLE (a) and SMLE (b).
	}
	\label{fig:ARE}
\end{figure}

We focus on the two-dimensional distributions first and then extend the simulations to three dimensions.\footnote{A Matlab code implementing the Bernstein-Kantorovich  sieve and  other codes used in this and the next sections are available at \url{http://research.economics.unsw.edu.au/vpanchenko/software/scopula.zip}.} Our simulation study is inspired by \cite{joe:05} who
studies the asymptotic  relative efficiency (ARE) of copula based MLE, i.e. the ratio of the inverse of asymptotic variance of QMLE to
that of FMLE.  He shows that
ARE depends on the specification of marginals and copula as well as on the strength of dependence. Moreover, for asymmetric marginal distributions, e.g., exponential, he finds that ARE for strongly negatively dependent data is much larger than for strongly positively dependent data holding the same absolute dependence strength.\footnote{Introducing a negative dependence in the DGP with two exponential marginals makes them skewed in the opposite directions. Accounting for the dependence in this case substantially helps with estimating the parameters of the marginals. We illustrate this later using the Fr\'{e}chet bounds.} We start our simulations from the bivariate DGPs with exponential marginals with distinct means $\mu_1=0.5,\mu_2=1$.\footnote{Note that it is easy to show analytically that for a multivariate distribution with exponential marginals and an arbitrary copula function, ARE of QMLE relative to FMLE does not depend on the parameters in the marginals. Our simulations suggest that the same holds for the SMLE. For generic marginals, ARE depends on both parameters in the marginals and the dependence parameter.} The dependence is modeled with 
bivariate 
Gaussian, Clayton, Plackett, and Frank copulas. Figure \ref{fig:ARE} reports AREs -- panel (a) FMLE vs QMLE and panel (b) SMLE vs QMLE -- as a function of dependence strength measured by Spearman's $\rho$ for the various copulas we use. Spearman's $\rho$ varies in the range $[-0.8, 0.8]$. Note that we use  the Clayton copula only for positive dependence.\footnote{The Clayton copula can be extended to incorporate negative dependence, but certain regions would have zero density \cite[][p.~158]{joe1997multivariate}.}
The SMLE asymptotic variance is estimated using (\ref{eq:asvar}) for a sample of 1,000,000 observations, where we use the tensor product sieve with cosine basis functions without the constant term to approximate $g_q$.  The number of sieve elements is $10\times10 = 100$.

Figure \ref{fig:ARE} confirms that there is a significant scope for improvement over the QMLE and that the largest gains are in the case of strong negative dependence. Naturally, the efficiency gains reported in Figure \ref{fig:ARE} using FMLE (panel a) are higher than those obtained using SMLE (panel b).  As expected the AREs of both FMLE and SMLE are near one (subject to some estimation noise) in the case of independence. 
We observe the lowest ARE, that is the biggest efficiency gains, when Spearman's $\rho$ approaches~$-1$. This corresponds to extreme negative dependence and agrees with 
\cite{joe:05}. In fact, SMLE asymptotic variance bounds  with copula parameters corresponding to Spearman's $\rho=-0.8$ suggest improvements of 31-54\% over the QMLE asymptotic variance depending on  copula. In the case of strong positive dependence FMLE does not show much efficiency gain over QMLE, which also agrees with \cite{joe:05}. The simulations summarized in  Figure \ref{fig:ARE} panel (b) show that similar patterns hold for SMLE. 

\citet[][Section 3]{joe:05} provides a detailed, identification-based, explanation for the asymmetry in ARE of FMLE with respect to the sign of $\rho$ by considering limiting dependence cases known as upper and lower Fr\'{e}chet bounds.  At a bound, there is an exact functional relation (different for the upper and lower bound) between the two dependent variables, $y_1=h(y_2;\beta_1,\beta_2)$. If parameters of the marginals $\beta_1$ and $\beta_2$ can be identified from this functional relationship, efficiency gains can be expected and this  happens for some asymmetric distribution families on the Fr\'{e}chet lower bound, and not for others.


\begin{figure}[!t]
	\centering
	\subfigure[]{\includegraphics[width=0.48\textwidth]{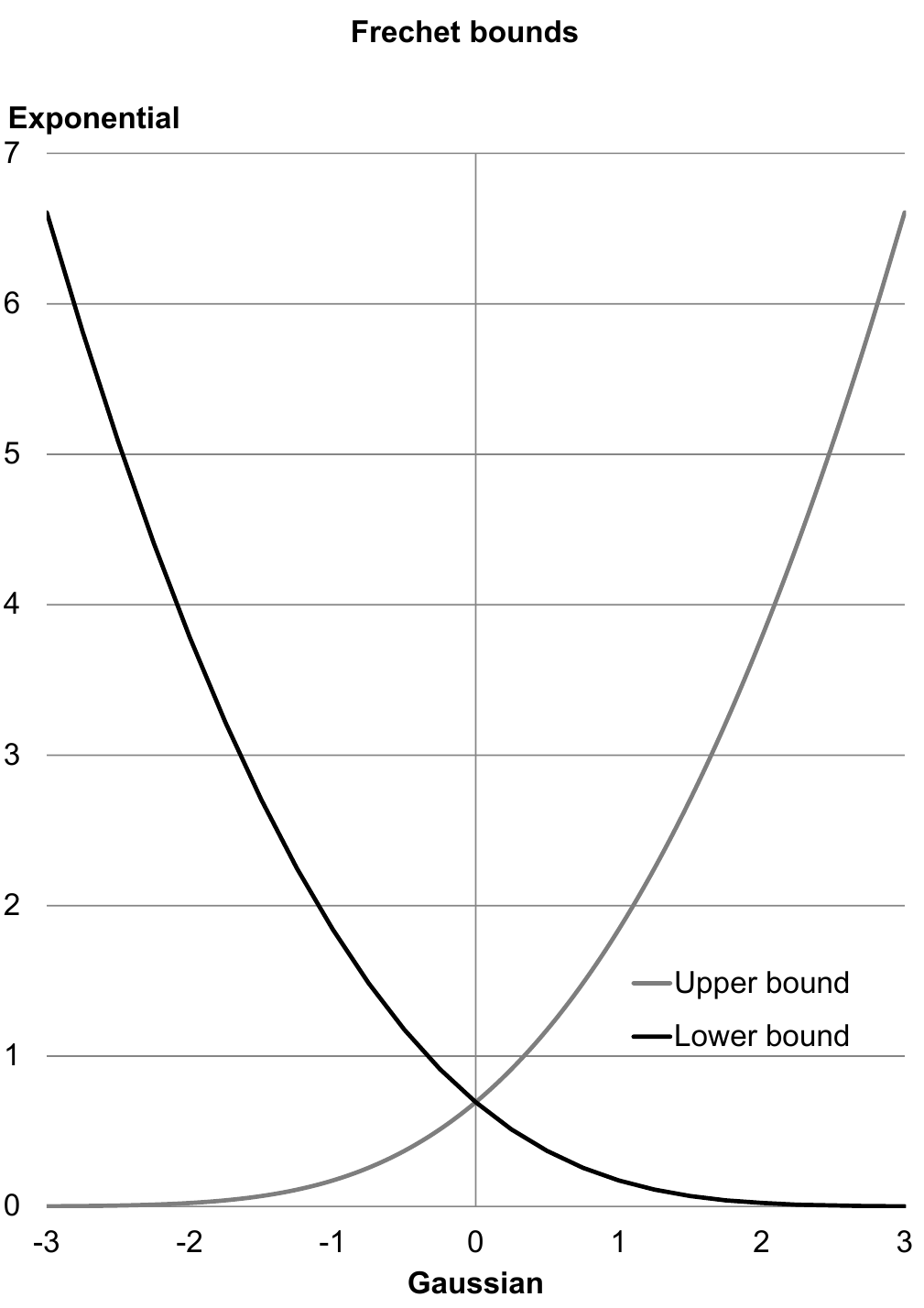}}\hfill
	\subfigure[]{\includegraphics[width=0.48\textwidth]{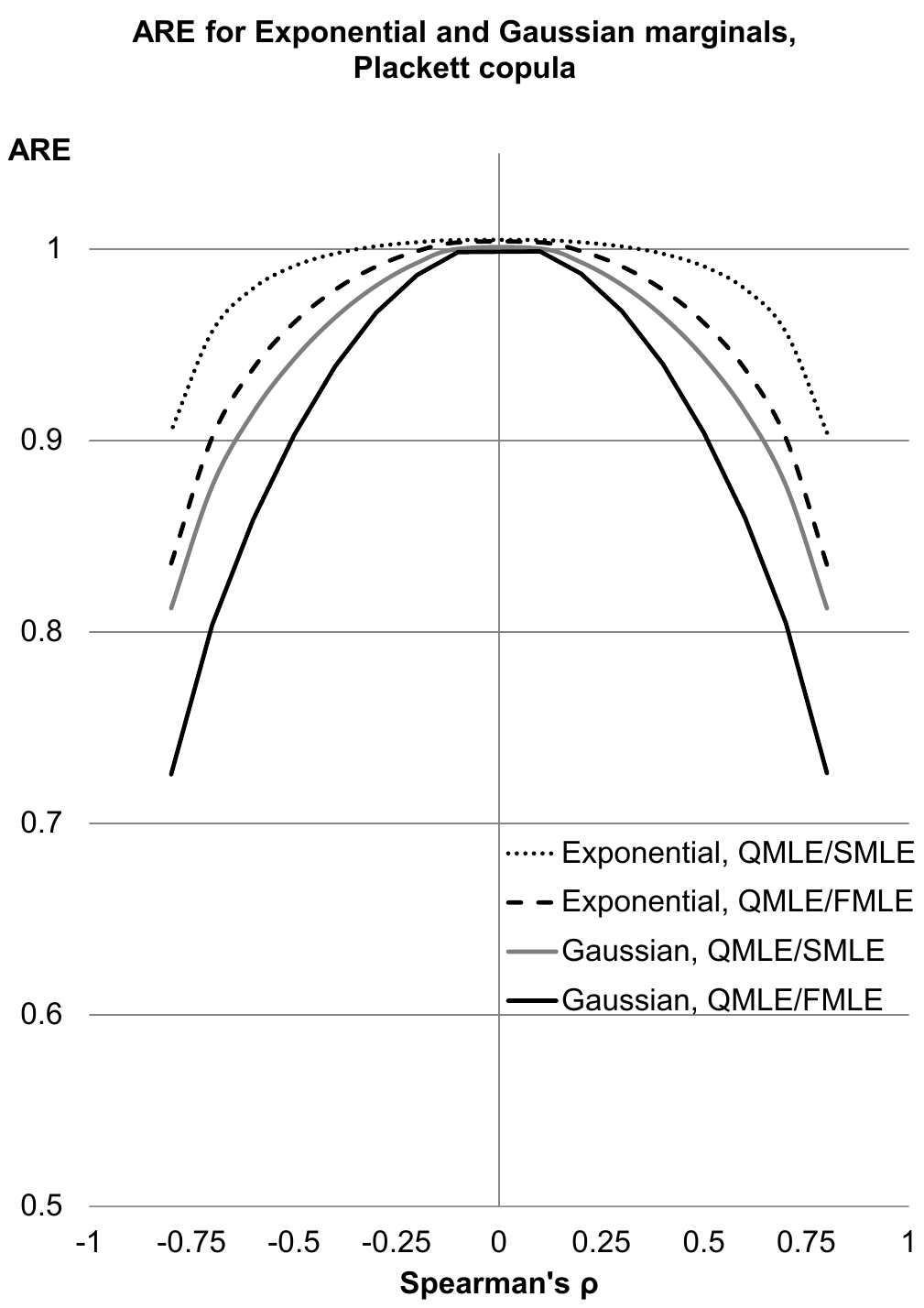}}
	\caption{Fr\'{e}chet bounds (a) and ARE of QMLE to FMLE and SMLE (b), symmetric case.
	}
	\label{fig:ARE-mixed}
\end{figure}

If any of the marginals is symmetric, the shape of the Fr\'{e}chet upper and lower bounds is the same 
and there is no difference in efficiency for the negative and positive side of the plot. 
We illustrate this by simulations with a bivariate DGP that has exponential and Gaussian marginals.\footnote{We ran a similar analysis for other marginal distributions, such as Gaussian with both mean and variance as unknown parameters (rather than fixed $\sigma^2$), skew-Gaussian, Pareto, logistic, gamma and some of their combinations (not reported for brevity). The general patters of the efficiency gains are similar to those reported in Figures 1 and 2.} Figure~\ref{fig:ARE-mixed}, panel~(a), shows the Fr\'{e}chet upper and lower bounds for the bivariate distribution that has an exponential marginal with $\mu_1=1$ and a Gaussian marginal with $\mu_2=0$ and $\sigma^2=1$ (assumed to be fixed). Panel (b) shows the AREs of FMLE to QMLE and the ARE of SMLE to QMLE for this DGP. 


Next we investigate in detail the performance of FMLE, SMLE and QMLE for a fixed value of Spearman's $\rho$. We use the exponential marginals and Plackett copula as the DGP and set the true parameter values in the marginals at $\mu_1=0.5$ and $\mu_2= 1$ and in the copula at $\gamma = 0.05$. We also include Pseudo MLE (PMLE), where the parametric functional form of the copula is misspecified. In addition, we considered a weighted score estimator, IQMLE of \cite{prokhorov/schmidt:09b}, but the results were essentially identical to QMLE and are not included (see \cite{prokhorov/schmidt:09b}, Theorem~1 (part b) for the explanation).  This illustrates the case when marginals are not restricted to have identical parameters and the dependence is moderate negative with Spearman's $\rho$ of $-0.77$. For the PMLE we use the Gaussian copula (G), which is a popular choice, and the Clayton rotated copula (C). We are interested in the Clayton copula as it has asymmetric tails, the 90 degree rotation is performed to model negative dependence. The sample size is $1,000$ and the number of simulations is $1,000$. 

\begin{table}[tp!]
	\begin{center}
		\begin{threeparttable}	
    \resizebox{1\textwidth}{!}{
			\begin{tabular}{l|ccccc|ccccc}
				&\multicolumn{5}{c|}{ $\mu_1=0.5$}&\multicolumn{5}{c}{$\mu_2=1$} \\
				&  FMLE  & SMLE  & QMLE &  PMLE(G) & PMLE(C) & FMLE &  SMLE & QMLE & PMLE(G) & PMLE(C) \\
				\hline
				\hline
    Mean  & 0.5004 & 0.4980 & 0.5001 & 0.5004 & 0.4945 & 0.9992 & 0.9996 & 0.9991 & 0.9988 & 1.0300 \\
    $N \times$ Var & 0.1724 & 0.1916 & 0.2635 & 0.2100 & 0.2260 & 0.6956 & 0.7679 & 1.0346 & 0.8399 & 0.9714 \\
    $N\times$ MSE & 0.1725 & 0.1956 & 0.2635 & 0.2101 & 0.2567 & 0.6962 & 0.7681 & 1.0354 & 0.8413 & 1.8708 \\
    $N\times$ AVar & 0.1582 & 0.1797 & 0.2500 & 0.1634 &  0.0816  & 0.6329 & 0.7193 & 1.0000 & 0.6518 & 1.0082 \\

\hline
			\end{tabular}
  }
		\end{threeparttable}
	\end{center}
 \caption{Simulated mean and variance for Plackett copula based FMLE, SMLE, QMLE, PMLE(G) and PMLE(C) assuming Gaussian and Clayton rotated copulas, respectively.}\label{tab:simul}
\end{table}

Table~\ref{tab:simul} contains the simulation results. We report the mean value of the estimates for each marginal as well as various versions of the variance estimator and MSE, scaled by the sample size $N$. Under Var, we report sample variance estimates while under AVar, we report estimates of the asymptotic variance obtained using a solution to (\ref{eq:asvar}).  The number of elements in the Bernstein-Kantorovich sieve in one dimension is $J_N=9$ and in total $9 \times 9=81$. This number minimizes the sum of mean-squared errors for both estimates (see Table~\ref{tab:CVsimul}).  A key feature of the table is that SMLE shows substantial improvement over QMLE as well as over two PMLEs. For the two misspecified copulas, while the use of the Gaussian copula only decreases efficiency, the use of the Clayton rotated copula leads to a visible bias in addition to a loss in efficiency. The bias is a result of the use of an asymmetric copula when the true GDP is symmetric \cite[see][for a discussion on copula robustness]{prokhorov/schmidt:09b}.  The sample variance of SMLE is close to its asymptotic variance bound. 

One of the practical problems we face in implementing SMLE is the choice of the order
of sieve $J_N$. While some asymptotic
results on the rate of convergence and its dependence on $J_N$ are
available, they are not informative in the finite sample
situation. The literature on sieves suggests using typical model
selection techniques, such as AIC and BIC, and we compare these criteria.

\begin{table}[tp!]
	\begin{center}
		\begin{threeparttable}
\setlength{\tabcolsep}{4pt}  
	\resizebox{1\textwidth}{!}{
	\begin{tabular}{r|rrrrrrr|rrr|r}
$J_N$	&	Mean1	&	Mean2	&	Var1	&	Var2	&	MSE1	&	MSE2& sumMSE &	LogL	&	AIC	&	BIC	&   run-time	\\
	\hline
	\hline
2     & 0.4906 & 0.9798 & 0.2334 & 0.9179 & 0.3209 & 1.3245 & 1.6455 & -1106.44 & 2218.87 & 2233.60 & 0.063 \\
3     & 0.4916 & 0.9818 & 0.2217 & 0.8745 & 0.2915 & 1.2055 & 1.4969 & -1007.75 & 2027.50 & 2056.95 & 0.139 \\
4     & 0.4942 & 0.9872 & 0.2128 & 0.8491 & 0.2460 & 1.0140 & 1.2600 & -948.54 & 1919.08 & 1973.07 & 0.315 \\
5     & 0.4959 & 0.9919 & 0.2062 & 0.8279 & 0.2228 & 0.8942 & 1.1170 & -909.48 & 1854.97 & 1943.31 & 0.803 \\
6     & 0.4965 & 0.9947 & 0.2029 & 0.8007 & 0.2152 & 0.8285 & 1.0436 & -881.91 & 1817.83 & 1950.34 & 1.851 \\
7     & 0.4970 & 0.9970 & 0.1945 & 0.7779 & 0.2037 & 0.7871 & 0.9908 & -861.85 & 1799.71 & 1986.20 & 4.037 \\
8     & 0.4975 & 0.9987 & 0.1916 & 0.7699 & 0.1977 & 0.7715 & 0.9692 & -847.12 & 1796.24 & 2046.53 & 7.968 \\
9     & 0.4980 & 0.9996 & 0.1916 & 0.7679 & 0.1956 & 0.7681 & 0.9637 & -835.55 & 1803.10 & 2127.01 & 15.232 \\
10    & 0.4983 & 0.9996 & 0.1938 & 0.7751 & 0.1966 & 0.7753 & 0.9719 & -826.27 & 1818.54 & 2225.89 & 23.170 \\
11    & 0.4988 & 0.9986 & 0.1910 & 0.7802 & 0.1926 & 0.7821 & 0.9747 & -818.71 & 1841.42 & 2342.01 & 37.999 \\
12    & 0.4988 & 0.9978 & 0.1950 & 0.7892 & 0.1965 & 0.7941 & 0.9905 & -812.66 & 1871.31 & 2474.96 & 60.428 \\
13    & 0.4986 & 0.9969 & 0.1957 & 0.7948 & 0.1978 & 0.8046 & 1.0024 & -807.70 & 1907.40 & 2623.94 & 91.228 \\
14    & 0.4984 & 0.9964 & 0.1969 & 0.7923 & 0.1994 & 0.8054 & 1.0048 & -803.60 & 1949.20 & 2788.43 & 136.661 \\
15    & 0.4982 & 0.9962 & 0.2052 & 0.8048 & 0.2084 & 0.8190 & 1.0275 & -800.07 & 1996.15 & 2967.88 & 227.359 \\

	\hline
			\end{tabular}
			}
		
		\end{threeparttable}
	\end{center}
 	\caption{Optimal number of sieve elements in SMLE}\label{tab:CVsimul}
\end{table}

In particular, we investigate how the SMLE estimates change with the number of sieve elements. Table~\ref{tab:CVsimul} reports means, variances and MSEs, scaled by $N$, for the two estimates as well as the value of log-likelihood and popular model selection criteria, AIC and BIC.   We also report average run-time in seconds for a specific $J_N$ per one sample of 1,000 observations.\footnote{The simulations were performed on the UNSW computational cluster using multiple CPU-cores and Matlab parallel computing. The run-time is the average per one CPU-core.}   
The value of log-likelihood, LogL, increases as sieve complexity grows, as expected. On average, BIC selects an under-parameterized model ($J_N=5$), whereas AIC selects $J_N=8$, which is close to  $J_N=9$ under which the smallest sum of MSEs is reached in the simulations. We also investigated $K$-fold cross-validation, but it was computationally expensive and did not provide any extra insights in addition to AIC. We note a degree of stability of SMLE for sufficiently high $J_N$, also noted by \citet[][Table 2]{sancetta:satchell:04}.

It is well known that nonparametric and semiparametric models are subject to the curse of dimensionality. Semiparametric copula sieve models are also affected by this issue. It also manifests itself in the non-linear relation between the run-time and the number of sieve elements $J_N$ (see the last column of Table \ref{tab:CVsimul}).
Nonetheless, there are SMLE efficiency gains possible even in three dimensions, as we demonstrate in the simulations below and in the application at the end of Section~\ref{sec:emp}. 
\begin{figure}[t!]
	\centering
	\subfigure[]{\includegraphics[width=0.48\textwidth]{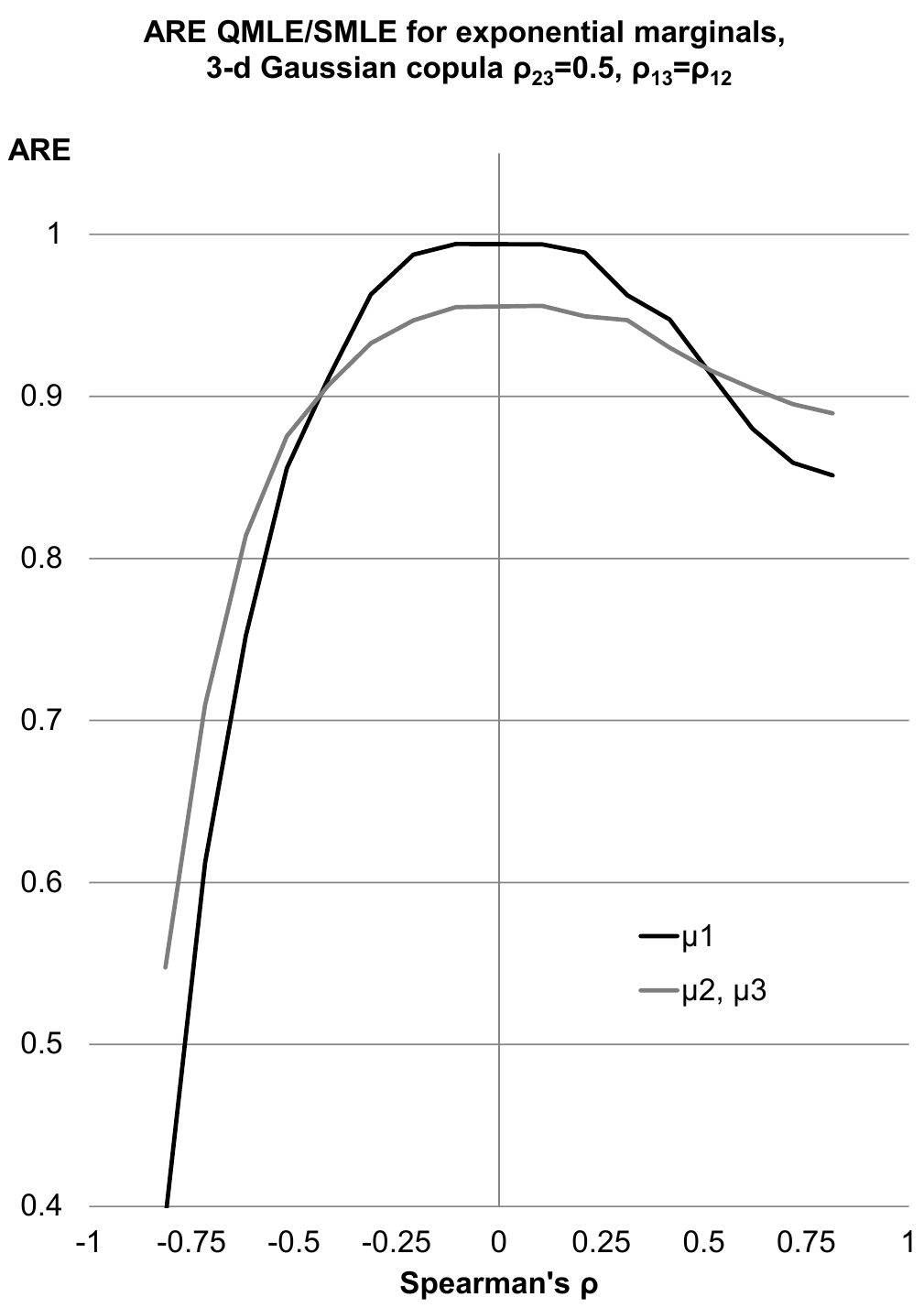}}\hfill
	\subfigure[]{\includegraphics[width=0.48\textwidth]{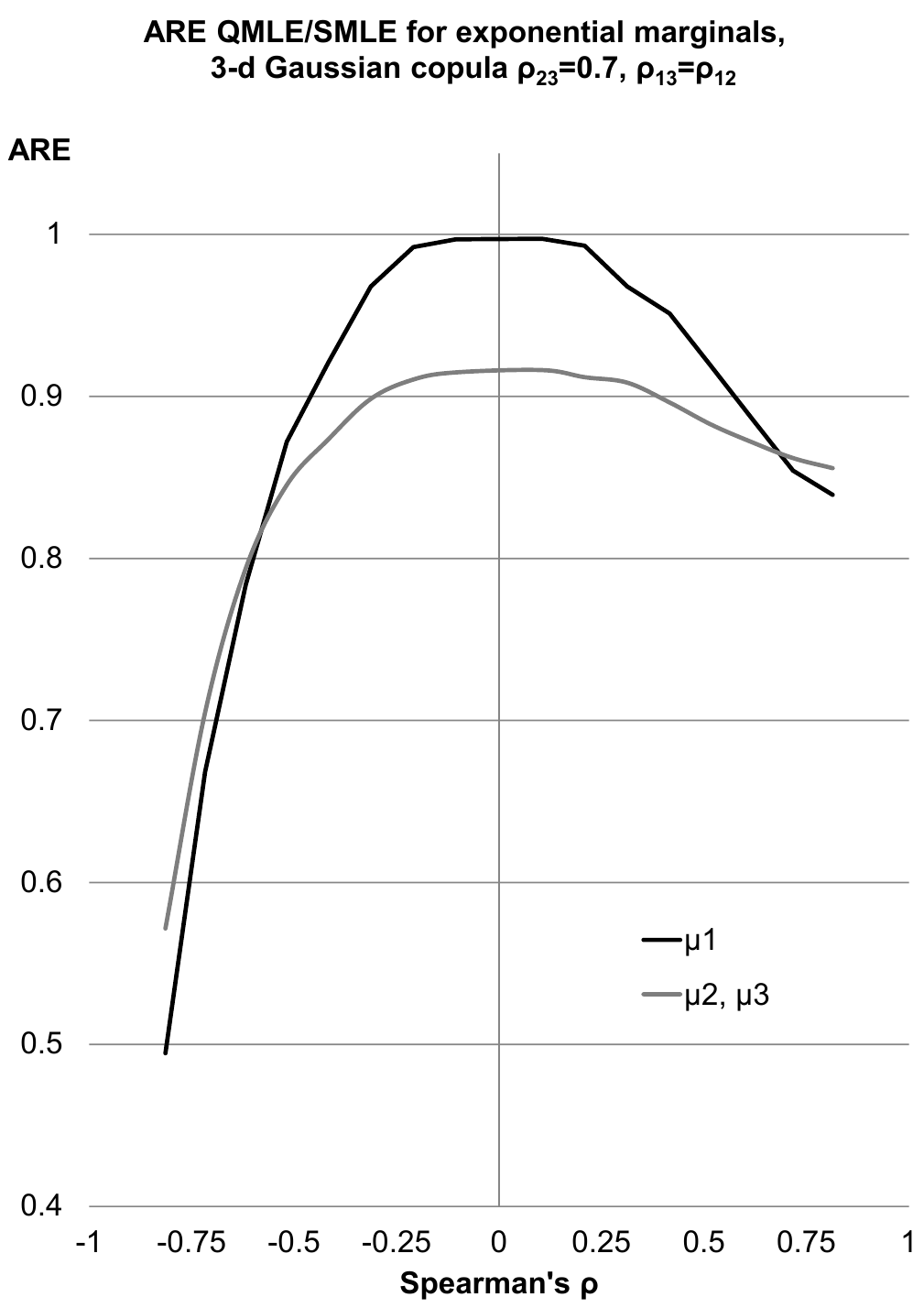}}
	\caption{ARE of SMLE for three dimensional copula.
	}
	\label{fig:ARE-3d}
\end{figure}

For easier comparison with the bivariate case we continue working with the exponential marginals labeled $1,2$ and $3$, with parameters $\mu_1=0.1$, $\mu_2=0.5$, and $\mu_3=1$, respectively. The dependence is specified with the Gaussian copula with three parameters, $\rho_{12}, \rho_{13}, \rho_{23}$, that reflect the dependence between the pairs of the marginals indicated by the subscripts. We vary $\rho_{12}$ and $\rho_{13}$ keeping them equal to each other and taking values between strong negative dependence and strong positive dependence on the scale analogous to Figures~\ref{fig:ARE} and \ref{fig:ARE-mixed}; 
the value of $\rho_{23}$ is fixed. Figure~\ref{fig:ARE-3d} compares ARE of SMLE relative to QMLE for two scenarios: panel (a) $\rho_{23}=0.5$ and panel (b) $\rho_{23}=0.7$.\footnote{There is a restriction on (negative) dependence range in the trivariate case, i.e., positive-definiteness of the corresponding correlation matrix. This motivates our choice for $\rho_{23}$ given the full range we consider for $\rho_{12}$ (and $\rho_{13}$). 
Since the marginals are exponential, as in the bivariate case, AREs do not depend on the value of the parameter in the marginals.} We observe that substantial efficiency gains can be realised in the trivariate case. AREs for $\mu_1$ on both panels are similar because these are mainly driven by $\rho_{12}=\rho_{13}$. AREs for $\mu_2$ and $\mu_3$ are essentially the same because of the assumed dependence structure. When comparing the ARE of SMLE between panel b and panel a, one can see that higher $\rho_{23}$ brings higher efficiency gains (especially, for $\mu_2$ and $\mu_3$) for the mid-range of $\rho_{12}=\rho_{13}$ but not near the extreme values of Spearman's $\rho$. 

\section{Empirical Application: Portfolio Value-at-Risk}
\label{sec:emp}

We consider an application from  investment portfolio management and show how the use of SMLE can lead to superior estimates of the market risk associated with an investment in a portfolio security such as a stock or bond.

Let $P_t$ denote the market price of a security at time $t$, and let $R_t$ be the associated holding period return between times $t$ and $t+1$. In addition to the rate of expected return $\mu_{R_t} = E[R_t]$ over the holding period, of key interest to an investor  are measures of riskiness  such as  variance  $\sigma^2_{R_t} = E[(R_t - E[R_t])^2]$ and ``Value-at-Risk'' (VaR), which, given some confidence level $\alpha \in [0,1]$, is defined as $V_{1-\alpha}(R_t) =\inf \{ R_t:F(R_t;\theta) > 1-\alpha \} = F^{-1}(1-\alpha;\theta)$, where $F$ is the c.d.f. of $R_t$ with parameter vector $\theta$. The $5\%$ VaR, or $V_{0.05}(R_t)$, for example, shows a level of loss that is only $5\%$ likely to be exceeded between times $t$ and $t+1$, and this measure is widely used in the financial industry to characterize ``worst-case'' scenarios associated with an investment.

A VaR estimate can be easily obtained as $\hat{V}_{1-\alpha}(R_t) = F^{-1}(1-\alpha;\hat{\theta})$, where $\hat{\theta}$ is the estimate of the marginal distribution parameter vector $\theta$. The VaR is therefore a functional of $\hat{\theta}$, and the properties of the estimator 
directly affect the accuracy of the density estimate $\hat{F}(R_t; \hat{\theta})$, and in turn the accuracy of the VaR measure. 
This represents an new avenue in VaR analysis since efficiency gains offered by SMLE 
of $\theta$ may translate into superior estimates of the VaR, without the need to specify the full joint distribution. That is, by using another variable associated with $R_t$, SMLE may produce 
a better estimate of the VaR for $R_t$, while avoiding the risk of biasing $\hat{V}_{1-\alpha}$ due to an incorrect choice of the dependence structure. 

\subsection{Estimating 5\% VaR for Bank of America Stock}

We begin by using SMLE to estimate weekly $5\%$ VaR for a potential investment in the Bank of America (NYSE: BAC) stock. To explore  improvements in the BAC VaR estimates arising from SMLE we need to find other variables that 
contain additional information about BAC returns. To this end, we select BAC trading volume and realized volatility of BAC returns. 
The return-volatility and return-volume relationships are both well-documented in the literature \cite[see, e.g.,]{gervais2001high, ang2006cross}. 
 
Our sample consists of daily adjusted closing prices for BAC beginning in August of 1999 and ending in December of 2020, and we use Friday closing price $P_t$ to calculate weekly holding period returns as $R_t = \ln(P_t / P_{t-1})$. For each of the weeks we also calculate the change in dollar trading volume during the week as $M_t = \ln(m_t / m_{t-1})$, where $m_t = \sum_j s_j P_j$, with $s_j$ representing the number of shares traded during the $j$-th day, with the summation being over all trading days of the week. We further estimate the weekly realized volatility $V_t$ of the BAC stock price as the standard deviation of daily returns during the week. 

We begin by constructing the marginal models. 
For $R_t$, we select Student-t, with density
 \begin{equation}
f_r(R_t; \nu_r,\sigma_r,\mu_r) = \frac{\Gamma \left(\frac{\nu_r+1}{2}\right)}{\Gamma \left( \frac{\nu_r}{2}\right) \sqrt{\pi\nu_r}\sigma_r} \left(1+\frac{1}{\nu_r}\left(\frac{R_t-\mu_r}{\sigma_r}\right)^2 \right)^{-\frac{\nu_r+1}{2}},
\label{eq:marginal_model_returns}
\end{equation} where $\Gamma(\cdot)$ is the Gamma function, $\mu_r$ and $\sigma_r$ are the location and scale parameters, and $\nu_r$ is the tail parameter which allows for excess kurtosis often present in financial returns. We adopt the same specification, with distinct parameters, for $M_t$ and denote that density  by $f_m(M_t; \nu_m,\sigma_m,\mu_m)$. For $V_t$, we chose the Beta distribution so that to capture non-negativity and skewness in volatility, with the density\begin{equation}
\label{eq::martinal_model_volatility}
    g(V_t; \alpha, \beta) = \frac{V_t^{\alpha - 1}(1-V_t)^{\beta-1}}{B(\alpha, \beta)}, \quad\text{where}\quad B = \frac{\Gamma(\alpha)\Gamma(\beta)}{\Gamma(\alpha + \beta)}.
\end{equation} 

The setup of Section \ref{section::sieve_mle} is general enough to accommodate any dimension, meaning that we can use SMLE to augment estimation of the marginal parameters $\mu_r$, $\sigma_r$, and $\nu_r$ using any number of auxiliary variables. 
To avoid the curse of dimensionality, here we restrict ourselves to two.  To better understand the relative improvements, 
we first add $V_t$ and $M_t$ separately. We build individual bivariate models for $(R_t, M_t)$ and $(R_t, V_t)$ in Section \ref{section::var_example_two_dimensions} and study the behaviour of the resulting VaR estimate before considering 
a trivariate model for $(R_t, M_t, V_t)$ in Section \ref{section::var_example_three_dimensions}. In all three cases, as benchmarks 
we use the QMLE under the assumption of independence between these variables, and 
the FMLE with a bivariate t-copula. In addition to correlation, the t-copula captures dependence in the tails of the joint distribution and is a  common choice in the financial industry today. Most importantly, we obtain a set of parameter estimates using SMLE. To compare the resulting sets of VaR measures we use likelihood-based scoring rules proposed by \cite{diks2011likelihood}.

\subsection{The case of two dimensions
}
\label{section::var_example_two_dimensions}

First, 
we estimate the marginal parameters in $F_r(R_t; \nu_r, \sigma_r, \mu_r)$ using QMLE under independence between BAC returns and trading volume, which is based on the log-density
\begin{equation}
    \ln h(R_t, M_t; \theta) = \ln f_r(R_t; \nu_r,\sigma_r,\mu_r) + \ln f_m(M_t; \nu_m,\sigma_m,\mu_m).
\end{equation}
This is equivalent to running two univariate MLEs of  the marginal parameters $(\nu_r,\sigma_r,\mu_r)$ and $(\nu_m,\sigma_m,\mu_m)$. We then re-estimate the parameter vector using FMLE, where we allow for a degree of correlation as well as tail dependence between BAC returns and changes in the trading volume by selecting the bi-variate t-copula in place of independence. The FMLE is based on 
\begin{eqnarray}
\label{eq::return_volume_gaussian_model}
\ln h(R_t, M_t; \theta) = \ln f_r(R_t; \nu_r,\sigma_r,\mu_r) + \ln f_m(M_t; \nu_m,\sigma_m,\mu_m) +c_t(F_r(R_t;\mu_r,\sigma_r,\nu_r),F_m(M_t;\mu_m,\sigma_m,\nu_m);\rho, \tau),
\end{eqnarray} where $c_t$ is the bivariate t-copula log-density parameterized by the correlation coefficient $\rho$ and tail thickness parameter $\tau$. In constructing the copula term in (\ref{eq::return_volume_gaussian_model}) we use the Bernstein-Kantorovich sieve from (\ref{eq:bernstein-kantorovich-sieve}) with $J_N = 5$ and estimate the marginal parameters in (\ref{eq:marginal_model_returns}) using SMLE. 

To construct the bivariate model for the case where returns are paired with volatility instead of trading volume we repeat these steps but replace the volume density $f_m$ in (\ref{eq::return_volume_gaussian_model}) with $g$ from (\ref{eq::martinal_model_volatility}).

We use a historical three-year rolling window to obtain all estimates and calculate corresponding weekly 5\% VaR for BAC for each of the trading weeks in the sample as $\hat{V}_{t, 0.05}(R_t) = \hat{F}_t^{-1}(0.05;\hat{\nu}_r,\hat{\sigma_r},\hat{\mu}_r)$, with the marginal parameters $(\hat{\nu}_r,\hat{\sigma_r},\hat{\mu}_r)$ obtained using QMLE, FMLE, and SMLE. 
To compare the behavior of these VaR measures and to gauge economic significance of the differences between them, 
we focus on two criteria: the number of times actual losses exceed the corresponding VaR estimate (which we refer to as exceedances), and the relative accuracy of the VaR, which we assess using the likelihood-based scoring rule of \cite{diks2011likelihood}.

VaR exceedances that occur more frequently than $(1-\alpha)\%$ of the time may suggest a biased VaR, and avoiding this bias is particularly important for institutional investors such as banks. Many larger banks are subject to capital adequacy requirements that are part of regulatory frameworks, for example the Bank of International Settlements Basel III Accord. Through their compliance process banks must maintain a portion of  capital invested in risk-free assets as security against possible trading losses. Such ``regulatory capital'' acts as a cushion against default in times of extreme market volatility and its size generally depends on the aggregate level of risk associated with the bank's balance sheet. VaR is often used as part of this risk calculation, and banks face financial penalties when the rate of exceedances is too high, and consequently the level of regulatory capital is too low.

We find the rate of exceedances for SMLE VaRs to be virtually the same as that for QMLE and FMLE, suggesting that any differences in the behavior of SMLE VaR are not due to the presence of bias in the VaR measure, and will not come at a negative economic cost to a would-be user.

The accuracy of our VaR forecasts largely depends on the accuracy of underlying density estimates $\hat{F}(R_t;\hat{\nu}_r,\hat{\sigma_r},\hat{\mu}_r)$ obtained using QMLE, FMLE and SMLE. We follow \cite{diks2011likelihood} and adopt a censored likelihood-based scoring rule for assessing the accuracy of competing density forecasts in the specified region of interest, which in our case is the left tail of the BAC return distribution. For each trading wees in the sample we calculate censored scores for QMLE, FMLE and SMLE-based density estimates as 
\[S(R_{t+1}; \hat{f}_r) \equiv w_t(R_{t+1})\ln \hat{f}_r(R_{t+1}; \hat{\mu}_r^t, \hat{\sigma}_r^t, \hat{\nu}_r^t) + (1-w_t(R_{t+1}))\ln \left ( 1 - \int w_t(s) \hat{f}_r^t(s; \hat{\mu}_r^t, \hat{\sigma}_r^t, \hat{\nu}_r^t) ds\right ), 
\] where $\hat{\mu}_r^t, \hat{\sigma}_r^t,$ and $\hat{\nu}_r^t$ are the marginal parameters of the return distribution which we estimate using a rolling window ending in period $t$. The function $w(s)$ weighs observations proportional to their distance from the left tail. In the extreme case $w(s)$ can be defined as an indicator, discarding all returns that fall above a certain threshold. We adopt a specification where $w(s) = 1/(1 + \exp(a(y-s))$, letting all observations  influence the score, while enabling higher weights for observations in the left tail. For the purpose of weighting we set the return threshold $y$ to negative $8\%$, which is the fifth percentile of $R_t$ in the whole sample. The parameter $a$ in $w(s)$ determines the rate with which weights diminish with distance from the threshold. We set $a = 30$, and also note that our results are not sensitive to alternative choices of these parameters, or the weighting function itself.

\begin{figure}
    \centering
    \includegraphics[scale=.9]{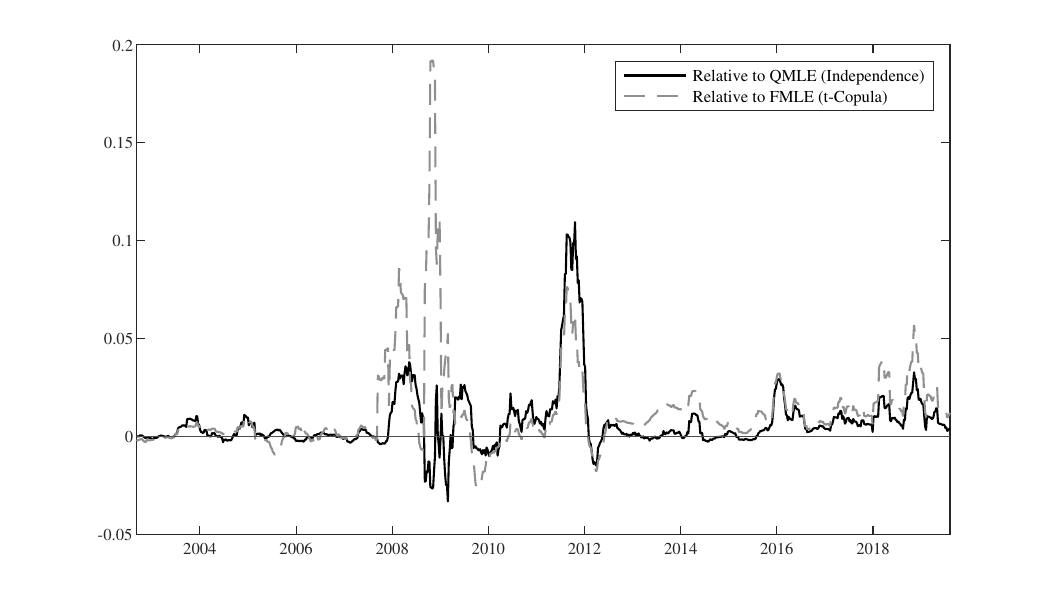}
    \caption{Differences in tail forecast accuracy between SMLE and QMLE,  and SMLE and FMLE. Twelve week moving average of score difference, January 2002 - December 2020.}
    \label{fig:smle_relative_accuracy_volume}
\end{figure}

Figure \ref{fig:smle_relative_accuracy_volume} shows smoothed differences in weekly values of $S(R_{t+1}; \hat{f}_r)$ for SMLE relative to QMLE and FMLE, for the case where returns are paired with volatility. Positive differences indicate higher SMLE scores, and for most of the trading weeks SMLE appears to produce significantly higher scores, meaning more accurate tail forecasts. 
Interestingly, this appears to be the case particularly in times of market turbulence, such as during the sub-prime crisis of 2007-2008 and the European debt crisis of 2011-2012. We also find this to be the case when returns are paired with trading volume $M_t$ instead of volatility.

\begin{table}[ht]
\begin{tabular}{lcccc}
                      & \multicolumn{2}{c}{Trading volume} & \multicolumn{2}{c}{Realized volatility} \\ \hline
                      & SMLE - QMLE       & SMLE - FMLE       & SMLE - QMLE          & SMLE - FMLE         \\ \hline
Mean difference & 0.0016                      & 0.0029                      & 0.0072                         & 0.0128                        \\
t-Ratio              & 2.2348                      & 4.0483                      &    3.1500                      & 5.6225  \\ \hline  
\end{tabular}
\caption{Differences in mean tail forecast accuracy scores between SMLE and QMLE, SMLE and FMLE.}
\label{table::returns_vol_volume_results}
\end{table}

To formally test for a positive difference between SMLE, QMLE and FMLE tail forecast scores we calculate differences in average scores in the sample and estimate the standard deviations of the differences using delete-$d$ jack-knife, with $d = 10$. Table \ref{table::returns_vol_volume_results} shows the results when returns are paired with realized volatility, and separately with trading volume. As before, a positive score difference indicates greater mean score for SMLE. In all cases, we reject the null hypothesis of equal mean scores in favor of a greater SMLE score at conventional significance levels.

Superior performance of SMLE-based VaR estimates is significant from a practical risk management standpoint and may be due to the SMLE's capability to better capture temporal shifts in market dependence as well as possible asymmetries, but we leave these questions for future work.

\subsection{The case of three dimensions}
\label{section::var_example_three_dimensions}

We now allows for simultaneous addition of volatility and trading volume. The QMLE in this case amounts to operating on the sum of marginal log-densities\begin{equation}
    \ln h(R_t, M_t, V_t; \theta) = \ln f_r(R_t; \nu_r,\sigma_r,\mu_r) + \ln f_m(M_t; \nu_m,\sigma_m,\mu_m) + \ln g(V_t; \alpha, \beta).
\end{equation} 
Using a trivariate t-copula $c_t(f_r, f_m, g; \Omega, \tau)$, parametrized by a correlation matrix $\Omega$ in addition to the tail thickness parameter $\tau$, yields the following log-density:
\begin{eqnarray}
\label{eq::log_likelihood_return_vol_volume}
\ln h(R_t, M_t; \theta) &=& \ln f_r(R_t; \nu_r,\sigma_r,\mu_r) + \ln f_m(M_t; \nu_m,\sigma_m,\mu_m) + \ln g(V_t; \alpha, \beta) \\\nonumber
&+& \ln c_t(F_r(R_t;\mu_r,\sigma_r,\nu_r),F_m(M_t;\mu_m,\sigma_m,\nu_m), G(V_t; \alpha, \beta);\Omega, \tau), 
\end{eqnarray}
where $ G(V_t; \alpha, \beta)$ is the cdf corresponding to $g(V_t; \alpha, \beta)$. As before, for SMLE we maximize the log-likelihood based on (\ref{eq::log_likelihood_return_vol_volume}) with respect to $\theta$, where the copula term is replaced with the Bernstein-Kantorovich sieve from (\ref{eq:bernstein-kantorovich-sieve}), now with $m = 3$, but keeping $J_N = 5$. 

We repeat all steps from the previous section and again follow \cite{diks2011likelihood} in obtaining forecast accuracy scores for SMLE, QMLE and FMLE, but we shorten the sample to 2012-2020 to accommodate increased computational complexity in higher dimensions.  Similar to the bivariate case, we find significant improvements in forecast accuracy in the tails when we used SMLE, with the differences in mean scores being positive and statistically significant at all conventional significance level. We summarize these results in Table~\ref{table::returns_vol_volume_together_results}.

\begin{table}[ht]
\begin{center}

\begin{tabular}{lcc}
\hline
 & \multicolumn{2}{c}{Trading volume and realized volatility} \\ \hline
 & Relative to QMLE & Relative to FMLE \\ \hline
SMLE score difference & 0.0022 & 0.0074 \\
t-Ratio & 2.3001 & 5.2096  \\ \hline
\end{tabular}
\caption{Differences in mean tail forecast accuracy scores between SMLE and QMLE, SMLE and FMLE.}
\label{table::returns_vol_volume_together_results}
\end{center}
\end{table}

\section{Concluding Remarks}

We have proposed an efficient semiparametric  estimator of
marginal distribution parameters.  This is a sieve maximum
likelihood estimator based on a finite-dimensional approximation
of the unspecified part of the joint distribution.  As such, the
estimator inherits the costs and benefits of the multivariate
sieve MLE. A major benefit is the
increased precision compared to quasi-MLE, permitted by the use of the dependence information. Simulations show
that potential efficiency gains are substantial. The efficiency bound is determined by the dependence strength and we show theoretically that our estimator reaches that bound.  
We illustrate the usability of SMLE with an empirical  applications in financial risk management (another application to insurance claims can be found in Appendix F). The dependence structure itself is not modeled directly which can be viewed as a drawback in some cases. However, the procedure has clear  advantages when the core interest is in estimating features of the marginals whereas dependence is viewed as a nuisance.  

The gains come at an increased computational expense.   The
convergence is slow for the traditional sieves we considered. 
We found that the Bernstein-Kantorovich polynomial is preferred to other sieves.  The running times are greater than the full MLE
assuming an ``off-the-shelf'' parametric copula family but far from being prohibitive
(at least in two and three-dimensional problems). 

In higher dimensions, the application of our approach is limited but a productive way to think about applying it is in the settings where one uses low dimensional copulas to arrive at a high-dimensional likelihood such as vine-copulas, factor copulas or composite densities \cite[see, e.g.,][]{scheffer/weiss:17, krupskii/joe:13, anatolyev/etal:18}. For example, \cite{scheffer/weiss:17} claim they were able to reach $d=15$ using vines of bivariate Bernstein copulas. We leave these methods for future work. 

Simple alternatives to the proposed method include a fully parametric
ML estimation problem and various weighting schemes of the QMLE moment conditions \cite[see, e.g.,][]{prokhorov/schmidt:09b, nikoloulopoulos/joe/chaganty:11}.  Although simpler computationally, the weighting schemes usually do not use information beyond correlation of the marginal scores. At the same time, the full MLE imposes correctness of the dependence structure, which, if
violated, renders the estimator inconsistent.  
Therefore, the proposed estimator seems to offer a natural  way of constructing a copula that is both robust and semiparametrically efficient. 

Methods to improve computational efficiency of SMLE focus on reducing the effective number of sieve parameters. Such methods involve penalized and restricted estimation and are particularly appealing for the Bernstein-Kantorovich polynomial where the sparse portions of the sieve parameter space correspond to histogram cells with little or no mass. We leave the development of such methods for future work. 

\section{Supplementary Material}

A Supplementary Materials file contains all the Appendices for this paper. An online package contains all Matlab codes implementing the Bernstein-Kantorovich sieve for the simulations and applications, and all the relevant data. The package can be found at  \url{http://research.economics.unsw.edu.au/vpanchenko/software/scopula.zip}.

\singlespacing
\bibliography{sieve1}
\doublespacing
\newpage

\end{document}